\documentstyle[preprint,eqsecnum,aps]{revtex}
\begin{document}
\draft
\title{Two Mode Quantum Systems: Invariant\\
Classification of Squeezing Transformations\\
and Squeezed States}
\author{Arvind\cite{email}}
\address{Department of Physics\\
Indian Institute of Science,  Bangalore - 560 012, India}
\author{B. Dutta}
\address{Jawaharlal Nehru Centre for Advanced Scientific Research\\
IISc Campus, Bangalore - 560 012, India}
\author{N. Mukunda\cite{jncasr}}
\address{Centre for Theoretical Studies and Department of Physics\\
Indian Institute of Science,  Bangalore - 560 012, India}
\author{and \\ R. Simon}
\address{Institute of Mathematical Sciences,
C. I. T. Campus, Madras - 600 113}
\date{\today}
\maketitle
\newpage
\begin{abstract}
A general analysis of squeezing transformations for two mode systems
is given based on the four dimensional real symplectic group
$Sp(4,\Re)\/$. Within the framework of the unitary metaplectic
representation of this group, a distinction between compact photon
number conserving and noncompact photon number nonconserving
squeezing transformations is made.  We exploit the
$Sp(4,\Re)-SO(3,2)\/$ local isomorphism and the $U(2)\/$ invariant
squeezing criterion to divide the set of all squeezing
transformations into a two parameter family of distinct equivalence
classes with representative elements chosen for each class. Familiar
two mode squeezing transformations in the literature are recognized
in our framework and seen to form a set of measure zero. Examples of
squeezed coherent and thermal states are worked out. The need to
extend the heterodyne detection scheme to encompass all of $U(2)\/$
is emphasized, and known experimental situations where all $U(2)\/$
elements can be reproduced are briefly described.
\end{abstract}
\newpage
\section{Introduction}
\label{introduction}
\setcounter{equation}{0}

The theoretical analysis~\cite{discovery} and
experimental~\cite{expt-one},
\cite{expt-two}, \cite{expt-three} realization of squeezed states of
radiation continue to receive a great deal of attention. While much
of the work so far has concerned itself with single mode
situations~\cite{reviews}, \cite{quadsq-one} some analysis of
two-mode states has also been presented~\cite{pcs},
\cite{pnd-two-mode}. Other nonclassical effects of radiation beyond
second order have also received attention in the
literature~\cite{high-order}. More recently, a
general invariant squeezing criterion for $n\/$-mode systems has been
developed by some of us elsewhere~\cite{multimode}.

The purpose of the present paper is to study squeezing
transformations for two-mode systems, and to develop a classification
scheme for them motivated by the above mentioned invariant squeezing
criterion.  Basic to all such discussions is the four-dimensional
real symplectic group $Sp(4,\Re)$, of real linear homogeneous
canonical transformations, and the unitary metaplectic representation
of this group acting on the Hilbert space of states for a two-mode
quantum system. The structure of the noncompact group $Sp(4,\Re)\/$ (
and its $n\/$-mode counterpart $Sp(2n,\Re)\/$) leads to a natural
separation of its elements into passive (compact), and active
(noncompact) types. Here the adjectives {\em passive\/} and {\em
active\/} mean total photon number conserving and nonconserving
respectively. This group theoretical framework gives us an
unambiguous way of defining precisely the family of squeezing
transformations; they are the active elements of $Sp(4,\Re)\/$ and
they do not form a subgroup. The action of the maximal compact
(passive) subgroup $U(2)\/$ of $Sp(4,\Re)\/$ on the set of squeezing
transformations by conjugation leads to a natural equivalence
relation, leading to the emergence of equivalence classes and
convenient representative elements as well. In studying the physical
properties of a state subjected to squeezing, therefore, we are able
to isolate the dependence on intrinsic squeezing parameters and
separate them from other passive factors.  As might be expected, the
single {\em squeeze factor\/} encountered in the studies of single
mode states gets enlarged here to {\em two independent intrinsic\/}
squeeze factors; and it turns out that the two mode squeezing
transformations so far studied in the literature form a very small
subset of all the independent available possibilities.

The material in this paper is arranged as follows: Section~\ref{two}
sets up the basic kinematics for two mode systems, the Fock and
Schr\"{o}dinger representations and the actions of $Sp(4,\Re)\/$ on
the canonical variables and states. The variance matrix for a general
state and its change under $Sp(4,\Re)\/$ are derived.  After
identifying the maximal compact or passive $U(2)\/$ subgroup of
$Sp(4,\Re)\/$, the $U(2)\/$ invariant squeezing criterion for two
mode systems is discussed. Section~\ref{three} introduces the
generators for the metaplectic representation of $Sp(4,\Re)\/$ and
brings out the connection to the $SO(3,2)\/$ Lie algebra. The photon
number conserving compact generators and the remaining noncompact
ones are clearly identified. Based on the polar decomposition theorem
for general elements of $Sp(4,\Re)\/$ we are then led to a precise
definition of squeezing transformations: These are single
exponentials of linear combinations of the noncompact metaplectic
generators. We then proceed to break up the set of all squeezing
transformations into equivalence classes under $U(2)\/$ action.  We
find that these classes form a continuous two parameter family
describable by points in an octant in a two dimensional plane.  The
families of Caves-Schumaker transformations and essentially single
mode transformations correspond to one dimensional lines bounding
this octant and so are of measure zero. Section~\ref{four} applies
our formalism to two mode squeezed coherent and thermal states. In
Section~\ref{five} we see how the heterodyne detection scheme fits
into our analysis. We argue that it is necessary to experimentally
realise all elements of the $U(2)\/$ subgroup of $Sp(4,\Re)\/$; the
heterodyne scheme only handles a one parameter subset of $U(2)\/$.
Two examples of two-mode situations where all elements of $U(2)\/$
can be experimentally realised are briefly described.
Section~\ref{concluding} contains some concluding remarks.

\section{Symplectic group for two modes and the squeezing criterion}
\label{two}
\setcounter{equation}{0}
We consider two orthogonal modes of the radiation field, with
annihilation operators $a_j, j=1,2$, and corresponding creation
operators $a^\dagger_j$.  These two modes could, for example, be two
different frequencies for the same or different propagation
directions and polarizations, two different propagation directions at
a common frequency, two different polarization states of plane waves
degenerate in frequency and direction of propagation, etc.  We
arrange these operators in the form of a four-component column
vector:
\begin{eqnarray}
\xi^{(c)} & = & (\xi^{(c)}_a) = \left( \begin{array}{c}
a_1 \\ a_2 \\ a^\dagger_1 \\ a^\dagger_2 \end{array}
\right),\nonumber\\
a & = & 1,2,3,4.
\end{eqnarray}
The superscript $(c)\/$ indicates that the entries here are complex,
i.e., nonhermitian, operators. For discussion of quadrature squeezing,
however, we need to also deal with the hermitian quadrature components
of these operators. Therefore we define another column vector $\xi\/$
with four hermitian entries, related to $\xi^{(c)}\/$ by a fixed
numerical matrix, as follows:
\begin{eqnarray}
\xi = (\xi_a) & = & \left( \begin{array}{c}
q_1 \\ q_2 \\ p_1 \\p_2 \end{array} \right), \nonumber \\
q_j=\frac{1}{\sqrt{2}}(a_j + a_j^\dagger) & , &
p_j = \frac{-i}{\sqrt{2}}(a_j - a_j^\dagger);\nonumber\\
\xi^{(c)} = \Omega \xi & , & \xi=\Omega^{-1}\;\xi^{(c)},
\nonumber \\
\Omega = (\Omega^{-1})^\dagger & = &  \frac{1}{\sqrt{2}}
\left(\begin{array}{cccc}
1&0&i&0\\0&1&0&i\\1&0&-i&0\\0&1&0&-i \end{array} \right).
\end{eqnarray}

The canonical commutation relations can now be written
either in terms of
$\xi\/$ or (in two ways) in terms of $\xi^{(c)}$:
\begin{eqnarray}
\mbox{}[\xi_a ,\xi_b ] & = & i\beta_{ab},\nonumber\\
\mbox{}[\xi^{(c)}_a ,\xi^{(c)}_b ] & = & \beta_{ab}, \nonumber\\
\mbox{}[\xi^{(c)}_a ,\xi^{(c)}_b{}^{\dagger}] & =
& (\Sigma_3)_{ab},\nonumber\\
(\beta_{ab}) & = & \left(\begin{array}{cccc}
0&0&1&0\\0&0&0&1\\-1&0&0&0\\0&-1&0&0 \end{array}\right),
\nonumber\\
\left((\Sigma_3)_{ab}\right)& =& \left(\begin{array}{cccc}
1&0&0&0\\0&1&0&0\\0&0&-1&0\\0&0&0&-1 \end{array}\right).
\label{commrel}
\end{eqnarray}

A general real linear homogeneous
transformation on the $q$'s and $p$'s is described
by a $4\times 4\/$
real matrix $S\/$ acting as follows:
\begin{eqnarray}
\xi  \rightarrow \xi^\prime & = & S\xi \nonumber \\
 \xi^\prime_a & = & S_{ab}\xi_b.
\label{sp4r-xi}
\end{eqnarray}
If the $\xi_a^\prime\/$ are to satisfy the same commutation relations
as the $\xi_a\/$, the condition on $S\/$ is:
\begin{equation}
S\; \beta\; S^T = \beta.
\label{symp-cond}
\end{equation}
This is the defining property for the elements of
the group $Sp(4,\Re)$,
which is a noncompact group:
\begin{equation}
Sp(4,\Re)=\left\{ S=4\times 4\; \mbox{real matrix}\;\Bigm\vert
S\; \beta\; S^T = \beta \right\}.
\label{sp4r}
\end{equation}
Note that $\beta = -\beta^{-1}\/$ itself is an
element of $Sp(4,\Re)\/$
whereas $\Sigma_3\/$ is not. Further, $S\in Sp(4,\Re)\/$ implies
$-S, S^T, S^{-1} = \beta\;S^T\beta^T \in Sp(4,\Re)\/$ and
$\det S = 1\/$
for every $S\in Sp(4,\Re)$.

The action of $Sp(4,\Re)\/$ on the nonhermitian
operators $\xi_a^{(c)}\/$
is by a (generally) complex matrix $S^{(c)}\/$ related to  $S\/$ by
conjugation with $\Omega$:
\begin{eqnarray}
S\in Sp(4,\Re):&\xi^\prime = S\xi\Rightarrow\nonumber\\
& \xi^{(c)}{}^\prime=S^{(c)}\; \xi^{(c)}, \nonumber \\
& S^{(c)} =\Omega \;S \;\Omega^{\dagger}.
\end{eqnarray}

We denote the Hilbert space on which $\xi\/$ and $\xi^{(c)}\/$ act by
${\cal H}$.  Since the hermiticity properties and commutation
relations of the $\xi_a\/$ are maintained by the transformation
(\ref{sp4r-xi}) for any $S\in Sp(4,\Re)$, and since the $\xi_a\/$ act
irreducibly on $\cal H\/$, it follows from the Stone-von Neumann
theorem~\cite{stone-neumann} that it should be possible to construct
a unitary operator ${\cal U}(S)\/$ on $\cal H\/$ implementing
(\ref{sp4r-xi}) via conjugation:
\begin{eqnarray}
S\in Sp(4,\Re): & S_{ab}\xi_b = {\cal U}(S)^{-1}\;\xi_a\;{\cal
U}(S)\,,
\nonumber\\
& {\cal U}(S)^\dagger\;{\cal U}(S) = {\bf 1}\;\mbox{on}\;{\cal H}\,.
\label{uopr-S}
\end{eqnarray}
This ${\cal U}(S)\/$, for each $S\/$, is clearly arbitrary upto (and
only upto by virtue of irreducibility) an $S\/$-dependent phase. Even
after making use of this freedom, however, it turns out that we
cannot choose the operators ${\cal U}(S)\/$ for various $S\/$ so as
to give us a true unitary representation of $Sp(4,\Re)\/$ on $\cal
H\/$. Rather, after maximum simplification, they give us a two-valued
representation of this group~\cite{metaplectic}:
\begin{equation}
S_1\,, S_2\in Sp(4,\Re): {\cal U}(S_1)\;{\cal U}(S_2) = \pm {\cal
U}(S_1\, S_2).
\end{equation}
Alternatively we can regard the ${\cal U}(S)$, chosen so as to obey
this composition law, as providing a true and faithful unitary
representation of the four-dimensional metaplectic group $Mp(4)$,
which in turn is a two-fold cover of $Sp(4,\Re)$. In the literature
the operators ${\cal U}(S)\/$ are often said to provide the
metaplectic representation of $Sp(4,\Re)$.  The connection between
$Mp(4)\/$ and $Sp(4,\Re)\/$ has some similarity with the one between
$SU(2)\/$ and $SO(3)\/$ familiar in angular momentum theory.

Now we consider physical states of the two-mode system, the action of
$Sp(4,\Re)\/$ on them, and the statement of a suitable squeezing
criterion.  Let $\rho\/$ be the density operator of any (pure or
mixed) state of the two-mode radiation field. With no loss of
generality we may assume that the means $\mbox{Tr}\,(\rho\xi_a)\/$ of
$\xi_a\/$ vanish in this state.  (Any non-zero values for these means
can always be reinstated by a suitable phase space displacement which
has no effect on the squeezing properties). Squeezing involves the
set of all second order noise moments of the quadrature operators
$q_j\/$ and $p_j\/$. To handle them collectively we define the
variance or noise matrix $V\/$ for the state $\rho\/$ as follows:
\begin{eqnarray}
V & = & (V_{ab}),\nonumber\\ V_{ab} & = & V_{ba} =
\frac{1}{2}\mbox{Tr}\,(\rho\{\xi_a,\xi_b\}).
\label{V-def}
\end{eqnarray}
This definition is valid for a system with any number of modes. For a
two-mode system it can be written explicitly in terms of $q_j\/$ and
$p_j\/$ as:
\begin{equation}
V=\left(\begin{array}{cccc}
\langle q_1^2\rangle&\langle q_1 q_2\rangle &
\frac{1}{2}\langle \{q_1,p_1\}\rangle &
\langle q_1 p_2\rangle \\
\langle q_1 q_2\rangle &\langle q_2^2\rangle &
\langle q_2 p_1\rangle &
\frac{1}{2}\langle \{q_2,p_2\}\rangle \\
\frac{1}{2}\langle \{q_1,p_1\}\rangle &
\langle q_2 p_1 \rangle &
\langle p_1^2\rangle& \langle p_1 p_2 \rangle \\
\langle q_1 p_2\rangle&
\frac{1}{2}\langle \{q_2,p_2\}\rangle&
\langle p_1 p_2 \rangle & \langle p_2^2 \rangle
\end{array} \right)
\label{V-expr}
\end{equation}
This matrix is real symmetric positive definite and obeys additional
inequalities expressing the Heisenberg uncertainty principle of
quantum mechanics~\cite{multimode}.

When the state $\rho\/$ is transformed to a new state $\rho^\prime\/$
by the unitary operator ${\cal U}(S)\/$ for some $S\in Sp(4,\Re)\/$,
we see easily from eqs (\ref{uopr-S}, \ref{V-def}) that the variance
matrix $V\/$ undergoes a symmetric symplectic transformation:
\begin{eqnarray}
S \in Sp(4,\Re):& \rho^{\prime} & ={\cal U}(S)\;\rho\;{\cal
U}(S)^{-1}
\Rightarrow\nonumber \\
& V^\prime & = S\;V\;S^T\,.
\label{vtransformation}
\end{eqnarray}
This transformation law for $V\/$ preserves all the properties
mentioned at eq.~(\ref{V-expr}).

Towards setting up a squeezing criterion, we identify an important
subgroup of $Sp(4,\Re)\/$, namely its maximal compact subgroup
$K\equiv U(2)\/$.  This consists of matrices $S\in Sp(4,\Re)\/$
having a specific block form determined by two-dimensional unitary
matrices belonging to $U(2)\/$:
\begin{eqnarray}
K = U(2) & = & \left\{S(X,Y) \in Sp(4,\Re)\Bigm\vert U = X - iY \in
U(2)
\right\}\subset Sp(4,\Re),\nonumber\\
S(X,Y) & = & \left(\begin{array}{cc} X & Y \\ -Y &
X\end{array}\right).
\end{eqnarray}
]Here $X\/$ and $Y\/$ are the real and imaginary parts of $U\/$; and
the unitary condition $U^\dagger\;U = {\bf 1}\/$ on $U\/$ ensures
that $S(X,Y)\/$ obeys the symplectic condition (\ref{symp-cond}). We
also recognise that this subgroup $K\/$ is the intersection of the
symplectic and the orthogonal groups in four real dimensions, namely
\begin{eqnarray}
\lefteqn{K = U(2) = Sp(4,\Re) \cap O(4):}\nonumber\\
& S(X,Y)\;(\beta\;\mbox{or}\;{\bf 1})\;S(X,Y)^T = \beta\;\mbox{or}\;
{\bf 1}.
\end{eqnarray}

The complex form $S^{(c)}(X,Y)\/$ corresponding to $S(X,Y)\/$, and
which acts on $\xi^{(c)}\/$, is rather simple; it is given by
\begin{eqnarray}
S^{(c)}(X,Y) & \equiv & S^{(c)}(U)\nonumber\\ & = &
\left(\begin{array}{cc} X-iY & 0 \\ 0 & X+iY\end{array}\right)
\end{eqnarray}
Thus the maximal compact subgroup of $Sp(4,\Re)\/$ mixes $a_1\/$ and
$a_2\/$ unitarily, {\em but does not mix the $a_j\/$ with the
$a_j^\dagger\/$}. This fact may be explained by saying that the
subgroup $U(2)\/$ of $Sp(4,\Re)\/$ consists of {\em passive\/}, or
total photon number conserving, transformations.

We see that the {\em physical requirement\/} that the total number of
photons be conserved singles out a unique maximal compact subgroup of
$Sp(4,R\/$) out of many equivalent ones.  In contrast, elements of
$Sp(4,\Re)\/$ outside the subgroup $U(2)\/$ are {\em non-compact\/}
elements which do {\em not\/} conserve total photon number, and so
describe {\em active\/} transformations. (These properties of {\em
compact\/} and {\em non-compact\/} elements of $Sp(4,\Re)\/$ will
become transparent when we identify their generators in the following
Section).

As has been discussed in detail elsewhere~\cite{multimode},
\cite{uone-sqcrt}, for a multimode system it is
physically reasonable to set up a definition of squeezing which is
invariant under the subgroup of passive transformations of the full
symplectic group. For the present case of two-mode systems, we
evidently need a $U(2)\/$-invariant squeezing criterion. That is, our
definition must be such that if a state $\rho\/$ with variance matrix
$V\/$ is found to be squeezed, then the state ${\cal
U}(S(X,Y))\;\rho\;{\cal U}(S(X,Y))^{-1}\/$ with variance matrix
$V^\prime = S(X,Y)\;V\;S(X,Y)^T$ must also be squeezed, for any $U =
X-iY\in U(2)\/$.

Conventionally a state is said to be squeezed if any one of the
diagonal elements of $V\/$ is less than 1/2. The diagonal elements
correspond, of course, to fluctuations in the ``chosen'' set of
quadrature components of the system. The $U(2)\/$-invariant
definition is as follows: the state $\rho\/$ is a quadrature squeezed
state if either some diagonal element of $V\/$ is less than 1/2 (and
then we say that the state is manifestly squeezed), or some diagonal
element of $V^\prime = S(X,Y)\;V\;S(X,Y)^T$ for some $U=X-iY\in
U(2)\/$ is less than 1/2:
\begin{eqnarray}
\lefteqn{\rho\;\mbox{is a squeezed state}\;\Leftrightarrow}\nonumber\\
&&\left(S(X,Y)\;V\;S(X,Y)^T\right)_{aa} < \frac{1}{2}\nonumber\\
&&\mbox{for some $a\/$ and some}\; X-iY\in U(2).
\label{sq-crt}
\end{eqnarray}
That is, running over $S(X,Y)\in K\/$ is the same as running over all
possible sets of quadrature components.  We may say that since any
element of $U(2)\/$ passively mixes the two modes, the appropriate
$S(X,Y)\in K\/$ which achieves the above inequality for some $a$
(assuming the given $V\/$ permits the same) just chooses the right
combination of quadratures to make the otherwise possibly hidden
squeezing manifest.

To implement this definition in practice, it would appear that even
if a state is intrinsically squeezed, we may have to explicitly find
a suitable $U(2)\/$ transformation which when applied to $V\/$ makes
the squeezing manifest. This however could be complicated. Here the
point to be noticed and appreciated is that diagonalisation of a
noise matrix $V\/$ generally requires a real orthogonal
transformation belonging to $SO(4)\/$ which may not lie in $U(2) =
O(4) \cap Sp(4,\Re)$. It is therefore remarkable that, as shown in
\cite{multimode}, the $U(2)\/$-invariant squeezing criterion
(\ref{sq-crt}) can be expressed in terms of the spectrum of
eigenvalues of $V\/$, namely:
\begin{equation}
\mbox{$\rho\/$ is a squeezed state}\;\Leftrightarrow\;\ell(V) =
\mbox{least eigenvalue of}\;V < \frac{1}{2}.
\label{sq-crt-mat}
\end{equation}
That is, while the diagonalization of $V\/$ is in general not
possible within $K = U(2)\/$ which is a proper subgroup of $O(4)\/$,
any one particular (and hence the smallest) eigenvalue of $V\/$ can
be made to become one of the diagonal elements of $V\/$ transformed
by an appropriate $S(X,Y) \in K$. In other words any quadrature
component can be taken to any other quadrature component by a
suitable element of $U(2)$. We shall in the sequel work with the
$U(2)\/$-invariant squeezing criterion (\ref{sq-crt},
\ref{sq-crt-mat}).

\section{The $Sp(4,\Re) - SO(3,2)\/$ connection and classification of
two-mode squeezing transformations}
\label{three}
\setcounter{equation}{0}
We have shown in the previous Section that the group $Sp(4,\Re)\/$ of
linear canonical transformations contains two kinds of elements:
passive total photon-number conserving elements belonging to the
maximal compact subgroup $K=U(2)$; and active {\em noncompact\/}
elements lying outside this subgroup, and which do not conserve total
photon number. It is clear from the $U(2)\/$-invariant squeezing
criterion (\ref{sq-crt},\ref{sq-crt-mat}) that the former elements
cannot produce squeezing. This is because the corresponding changes
in the variance matrix, $V^\prime = S(X,Y)\;V\;S(X,Y)^T$, being
similarity transformations preserve the eigenvalue spectrum of $V\/$;
hence $\ell(V)\geq 1/2\/$ implies $\ell(V^\prime)\geq 1/2\/$ and
conversely for every $S(X,Y) \in U(2)$.  The {\em noncompact\/}
elements of $Sp(4,\Re)\/$, on the other hand, while they do not form
a subgroup, have the potential to produce a squeezed state starting
from a nonsqueezed state. Thus they may be called squeezing
transformations. The following questions then naturally arise: what
are the really distinct squeezing transformations which are not
related to each other by just passive transformations, and how can
they be invariantly labelled or parametrised?

To answer these questions it is useful to work at the level of the
Lie algebra $\underline{Sp(4,\Re)}\/$ of $Sp(4,\Re)\/$, and the
hermitian generators of the unitary metaplectic representation ${\cal
U}(S)\/$ acting on $\cal H\/$. As is well known, these generators are
basically all possible hermitian symmetric quadratic expressions in
the canonical variables $q\/$ and $p$. They may be expressed more
transparently for our purposes in terms of $a$'s and $a^\dagger$'s.
While it is possible to set up the generators in a uniform manner for
all $Sp(2n,\Re)\/$, in the two-mode case we can exploit the fact that
the group $Sp(4,\Re)\/$ is locally isomorphic to the de Sitter group
$SO(3,2)\/$~\cite{sp4r-so32}, as they share the same Lie algebra. We
therefore choose the basis for $\underline{Sp(4,\Re)}\/$ to make this
explicit; it helps us visualise geometrically the analysis to follow
\cite{gilmore}.
We define the ten generators $Q$, $J_r$, $K_r$, $L_r$, $r=1,2,3\/$ of
$Sp(4,R)$ as:
\begin{mathletters}
\label{sp4r-gen}
\begin{eqnarray}
Q &=&
\frac{1}{2}(N+1)
= \frac{1}{2}(a_1^\dagger a_1 + a_2^\dagger a_2 + 1);\\ J_1 &=&
\frac{1}{2}(a^{\dagger}_1 a_2
+ a^{\dagger}_2 a_1), \nonumber \\ J_2 &=&
\frac{i}{2}(a^{\dagger}_2 a_1 -
a^{\dagger}_1 a_2), \nonumber \\ J_3 &=&
\frac{1}{2} (a^{\dagger}_1a_1 - a^{\dagger}_2a_2); \\
K_1 &=&
\frac{1}{4}(a^{\dagger}_1{}^{2} + a_1^2 -
a_2^{\dagger}{}^2 - a_2^2),\nonumber\\ K_2 &=& -
\frac{i}{4}(a_1^{\dagger}{}^2 - a_1^2 + a_2^{\dagger}{}^2
- a_2^2 ),\nonumber \\ K_3 &=& -\frac{1}{2}
(a^{\dagger}_1a^{\dagger}_2 +a_1a_2);\\ L_1 &=&
\frac{i}{4}(a_1^{\dagger}{}^2
- a_1^2 - a_2^{\dagger}{}^2 + a_2^2 ), \nonumber \\ L_2 &=&
\frac{1}{4}(a_1^{\dagger}{}^2 + a_1^2 +
a_2^{\dagger}{}^2 + a_2^2 ),\nonumber\\ L_3 &=& -\frac{i}{2}
(a^{\dagger}_1 a^{\dagger}_2 - a_1a_2).
\end{eqnarray}
\end{mathletters}
These particular quadratic expressions in $a_j$ and $a_{j}^{\dagger}$
are chosen because the commutation relations have a suggestive simple
form:
\begin{mathletters}
\label{comm-rel}
\begin{eqnarray}
\mbox{}[J_r,J_s] & = & i\epsilon_{rst}J_t,\nonumber\\
\mbox{}[Q,J_r] & = & 0;\label{u2-comm}\\
\mbox{}[J_r,K_s\;\mbox{or}\;L_s] & = & i\epsilon_{rst}(K_t
\;\mbox{or}\;L_t),\nonumber\\
\mbox{}[Q,K_r \pm iL_r] & = & \mp (K_r \pm iL_r);
\label{u2-ncmp-comm}\\
\mbox{}[K_r,K_s] & = & [L_r,L_s] = -i\epsilon_{rst}J_t,\nonumber\\
\mbox{}[K_r,L_s] & = & i\delta_{rs}Q.
\label{ncmp-comm}
\end{eqnarray}
\end{mathletters}
Comparing and combining the above two sets of equations we see that
the four operators $Q\/$ and $J_r\/$ are the photon-number conserving
generators of $U(2)\/$ (being respectively the generators of the
$U(1)\/$ and $SU(2)\/$ parts of $U(2)\/$); while the six operators
$K_r\/$ and $L_r\/$ are the {\em noncompact\/} generators of
squeezing transformations.

The connection to $SO(3,2)\/$ becomes evident by regarding these
generators as the various components of an antisymmetric set $M_{AB}
= -M_{BA}\/$, where the indices $A,B\/$ go over the range $1,2,\ldots
,5$:
\begin{eqnarray}
Q = M_{45}, & \qquad J_r & = \frac{1}{2}\epsilon_{rst}M_{st};
\nonumber \\
K_r = M_{r4}, & \qquad L_r & = M_{r5}.
\end{eqnarray}
Then the commutation relations (\ref{comm-rel}) take on the de Sitter
form
\begin{eqnarray}
[M_{AB}, M_{CD}] & = & i(g_{AC} M_{BD} - g_{BC} M_{AD} + g_{AD}
M_{CB} - g_{BD} M_{CA}),\nonumber\\ g_{AB} & = & \mbox{diag}\,.
(+1,+1,+1,-1,-1).
\end{eqnarray}
In this picture, $Q, J_r\/$ are generators of rotations in the 4-5
plane and in the 1-2-3 subspace respectively, together making up the
maximal compact $SO(2)\times SO(3)\/$ subgroup of $SO(3,2)$.  With
respect to the Fock or two mode photon number basis $\vert n_1
,n_2\rangle\/$ for $\cal H\/$ the situation is that all these vectors
with a fixed total number $n=n_1 + n_2\/$, $(n+1)\/$ in all, are
eigenvectors of $Q\/$ with a common eigenvalue $(n+1)/2\/$; and they
simultaneously provide the {\em spin\/} $n/2\/$ representation of the
$SO(3)\/$ subgroup generated by $J_r\/$. On the other hand, the
$K_r\/$ and $L_r, r=1,2,3\/$ are noncompact {\em Lorentz boost\/}
generators in the $r-4\/$ and $r-5\/$ planes respectively.

With this algebraic background, we go on to consider two-mode
squeezing transformations. It is a well known
fact~\cite{gilmore}\cite{siegel} that each matrix $S\in Sp(4,\Re)\/$
can be decomposed, globally and uniquely, into the product of two
particular kinds of $Sp(4,\Re)\/$ matrices: one factor belongs to the
subgroup $K$, the other to a subset $\Pi\/$ defined in the following
way:
\begin{equation}
\Pi = \left\{S\in Sp(4,\Re)\Bigm\vert S^T = S = \mbox{positive definite}
\right\}\subset Sp(4,\Re).
\end{equation}
We shall hereafter denote elements in $\Pi\/$ by
$P,P^\prime,\ldots\/$.  The decomposition mentioned above is then:
\begin{eqnarray}
S\in Sp(4,\Re)\;:\quad S & = & P\;S(X,Y),\nonumber\\ P & \in &
\Pi,\nonumber\\
X - iY & \in & U(2),
\end{eqnarray}
with both factors being uniquely determined by $S$. This is the form
in the present context of the familiar polar decomposition
formula~\cite{polar-decomposition} for a general matrix. For the
metaplectic operators ${\cal U}(S)\/$ we have the corresponding
statement
\begin{eqnarray}
{\cal U}(S) & = & {\cal U}(P)\;{\cal U}(S(X,Y)),\nonumber\\ {\cal
U}(P) & = & \exp\left[i(\mbox{real linear combination of }\;
\vec{K}\;\mbox{and}\;\vec{L})\right],\nonumber\\
{\cal U}(S(X,Y)) & = & \exp\left[i(\mbox{real linear combination of
}\; Q\;\mbox{and}\;\vec{J})\right].
\label{decomposition}
\end{eqnarray}

We may now identify precisely the most general squeezing
transformation within the $Sp(4,\Re)\/$ framework, as the operator
${\cal U}(P)\/$ characterised by two numerical three-dimensional
vectors $\vec{k}\/$, $\vec{l}\/$ appearing as coefficients of
$\vec{K}\/$ and $\vec{L}\/$ in the exponent:
\begin{equation}
{\cal U}(\vec{k},\vec{l}) = \exp\left[i(\vec{k}\cdot\vec{K} +
\vec{l}\cdot\vec{L})\right].
\label{utopr-so32}
\end{equation}
Thus we reserve the name {\em squeezing transformations\/} for these
elements of $\Pi\/$ within $Sp(4,\Re)\/$, represented in the
metaplectic representation by a single exponential factor. (It is
well to keep in mind that $\Pi\/$ is not a subgroup, so the product
of two such single exponential squeezing transformations is in
general not a similar single exponential. This is analogous to the
well known fact that the product of two $SO(3,1)\/$ Lorentz boosts is
not just a boost but a boost followed or preceded by a rotation
called the Wigner rotation~\cite{wigner}).

We may relate the decomposition (\ref{decomposition}) of a general
metaplectic transformation to a general quadratic Hamiltonian quite
directly.  Any such Hamiltonian containing both photon conserving and
nonconserving terms with possibly time dependent coefficients would
lead via the Schr\"{o}dinger equation to a unitary finite time
evolution operator which can be uniquely decomposed into the product
form (\ref{decomposition}).  Thus, integration of the Schr\"{o}dinger
equation leads in general to a specific passive factor and another
specific squeezing transformation.  In case the Hamiltonian is time
independent and a combination only of the generators $\vec{K}\/$ and
$\vec{L}\/$, this evolution operator is already of the form ${\cal
U}(P)\/$.

Since we have a $U(2)\/$-invariant squeezing criterion, as we have
seen, elements of $U(2)\/$ have no effect on the squeezed or
nonsqueezed status of any given state. This means that the $U(2)\/$
transform, by conjugation, of a squeezing transformation is another
squeezing transformation which should be regarded as equivalent to
the first one. It is clear that, in any case, any equivalence
relation among squeezing transformations as defined by us above
should be based on processes which take one squeezing transformation
to another.

Now from the commutation relations (\ref{u2-ncmp-comm}) we can see
that the generators $\vec{K}\/$, $\vec{L}\/$, and the squeezing
transformations ${\cal U}(\vec{k},\vec{l})\/$ defined in
eq.~(\ref{utopr-so32}), behave as follows under conjugation by
elements of $U(2)\/$:
\begin{mathletters}
\label{u2-ncmp}
\begin{eqnarray}
e^{\textstyle i\theta Q} \left( \vec{K}\pm i\vec{L}\right)
e^{\textstyle -i\theta Q}& =& e^{\textstyle \mp i\theta}
\left( \vec{K}\pm i\vec{L}\right),\nonumber\\
e^{\textstyle i\vec{\alpha}\cdot\vec{J}} \left( K_r\;\mbox{or}\; L_r
\right) e^{\textstyle -i\vec{\alpha}\cdot\vec{J}}& =&
R_{sr}(\vec{\alpha}) \left( K_s\;\mbox{or}\; L_s\right);
\label{u1:su2-ncmp}\\
e^{\textstyle i\theta Q}{\cal U}(\vec{k},\vec{l}) e^{\textstyle
-i\theta Q} & = & {\cal U}(\vec{k}^\prime,\vec{l}^\prime),\nonumber\\
\left(\begin{array}{c} \vec{k}^\prime \\ \vec{l}^\prime\end{array}\right)
& = & \left(\begin{array}{lr} \cos\theta & -\sin\theta\\
\sin\theta & \cos\theta\end{array}\right)
\left(\begin{array}{c} \vec{k}\\ \vec{l}\end{array}\right);
\label{u1-utopr}\\
e^{\textstyle i\vec{\alpha}\cdot\vec{J}} {\cal U}(\vec{k},\vec{l})
e^{\textstyle -i\vec{\alpha}\cdot\vec{J}}& =& {\cal
U}(\vec{k}^{\prime\prime},\vec{l}^{\prime\prime}),\nonumber\\
k_r^{\prime\prime}\;\mbox{or}\; l_r^{\prime\prime} & = &
R_{rs}(\vec{\alpha}) \left(k_s\;\mbox{or}\;l_s\right);
\label{su2-utopr}\\
R_{rs}(\vec{\alpha}) & = & \delta_{rs}\cos\alpha + \alpha_r
\alpha_s \frac{1 - \cos\alpha}{\alpha^2} + \epsilon_{rst}\alpha_t
\frac{\sin\alpha}{\alpha},\nonumber\\
\alpha & = & |\vec{\alpha}|.
\end{eqnarray}
\end{mathletters}
Here we have listed separately the effects of $U(1)\/$ and $SU(2)\/$
within $U(2)\/$ on the generators and the squeezing transformations.
The questions raised at the start of this Section can now be posed
more precisely: If the set of squeezing transformations
$U(\vec{k},\vec{l})\/$ is separated into distinct nonoverlapping
equivalence classes based on the $U(2)\/$ action (\ref{u2-ncmp}), how
can we conveniently choose $U(2)\/$-invariant parameters to label
these classes, and then pick a convenient representative element from
each class? We answer these questions in this sequence.

It is clear that we need to construct a complete set of independent
expressions in $\vec{k}\/$ and $\vec{l}\/$, invariant under both
$U(1)\/$ and $SU(2)\/$ actions (\ref{u1-utopr}, \ref{su2-utopr}).  We
begin by defining the matrix $M\/$ of scalar products among
$\vec{k}\/$ and $\vec{l}\/$, which is then $SU(2)\/$ invariant:
\begin{equation}
M(\vec{k},\vec{l})=\left(\begin{array}{cc}
\vec{k}\cdot\vec{k}&\vec{k}\cdot\vec{l}\\
\vec{k}\cdot\vec{l}&\vec{l}\cdot\vec{l}\end{array}\right).
\end{equation}
This is a real, symmetric positive semi-definite matrix. With respect
to $U(1)\/$ action, we see from eq. (\ref{u1-utopr}) that
$M(\vec{k},\vec{l})\/$ undergoes a similarity transformation by the
rotation matrix of angle $\theta\/$:
\begin{eqnarray}
M(\vec{k}^{\prime},\vec{l}^{\prime}) & = &
R(\theta)\;M(\vec{k},\vec{l})
\; R(\theta)^{-1}, \nonumber\\
R(\theta) & = & \left(\begin{array}{lr} \cos\theta & -\sin\theta\\
\sin\theta & \cos\theta\end{array}\right).
\label{uone-M}
\end{eqnarray}
One now sees that there are two independent $U(2)\/$ invariants that
can be formed:
\begin{eqnarray}
\Im_1(\vec{k},\vec{l}) = & \mbox{det\,}M(\vec{k},\vec{l}) & =
\vert\vec{k}\wedge\vec{l}\vert^2,\nonumber\\
\Im_2(\vec{k},\vec{l}) = & \mbox{Tr\,}M(\vec{k},\vec{l}) & =
\vert\vec{k}\vert^2 + \vert\vec{l}\vert^2;
\label{invts-gen}
\end{eqnarray}
and it is easily checked that there are no other invariants
independent of these.

Next let us tackle the problems of finding convenient parameters and
representative squeezing transformations for the $U(2)$ equivalence
classes, one for each class. We see from eq. (\ref{invts-gen}) that
if $\Im_1>0\/$ (i.e. $\Im_1\neq 0\/$) then the two vectors
$\vec{k}\/$ and $\vec{l}\/$ are both nonzero and nonparallel; while
if $\Im_1=0\/$ they are parallel (and one of them could vanish).
These are therefore clearly different geometrical situations.
Starting with the matrix $M(\vec{k},
\vec{l})\/$, we see from its $U(1)\/$ transformation law (\ref{uone-M})
that by a suitable choice of the angle $\theta\/$ we can arrange the
transformed matrix $M(\vec{k}^\prime,
\vec{l}^\prime)\/$ to be diagonal, and in the case of unequal
eigenvalues to place the larger eigenvalue in the first position.
This means that in each equivalence class of squeezing
transformations there certainly are elements ${\cal
U}(\vec{k},\vec{l})\/$ for which $\vec{k}\cdot\vec{l} = 0\/$ and
$\vert \vec{k}\vert \geq \vert \vec{l}\vert$. This still leaves us
the freedom of action by $SU(2)\/$. We may now exploit this freedom
to put the (mutually perpendicular) vectors $\vec{k}\/$ and
$\vec{l}\/$ into a convenient geometrical configuration.  A look at
the forms of the noncompact generators $\vec{K}\/$ and $\vec{L}\/$ in
eq. (\ref{sp4r-gen}) suggests that we choose $\vec{k}\/$ and
$\vec{l}\/$ as follows:
\begin{equation}
\vec{k} = (0,a,0),\quad \vec{l} = (b,0,0),\quad a \geq b.
\label{kl-simple}
\end{equation}
(A further reason for this choice will emerge shortly). $\Im_1\/$ and
$\Im_2\/$ can now be evaluated in terms of $a,b\/$ to obtain the
relations
\begin{eqnarray}
\Im_1(\vec{k},\vec{l}) & = & a^2 b^2,\nonumber\\
\Im_2(\vec{k},\vec{l}) & = & a^2  + b^2,\nonumber\\
a \geq b\geq 0& ,& (a,b)\neq (0,0).
\label{invts-sp}
\end{eqnarray}
With this parametrisation we can now say: there is a two-fold
infinity of distinct equivalence classes of squeezing transformations
for two-mode systems, each class corresponding uniquely and
unambiguously to a point $(a,b)\/$ in the octant $a\geq b\geq 0\/$ in
the real $a-b\/$ plane, excluding the origin.  Different points in
the octant correspond to intrinsically distinct equivalence classes.
Within an equivalence class determined by a point $(a,b)\/$, of
course, one can connect different squeezing transformations ${\cal
U}(\vec{k},\vec{l})\/$ by conjugation with suitable $U(2)\/$
elements. Given a squeezing transformation $P\in \Pi\subset
Sp(4,\Re)\/$ its class $(a,b)\/$ is determined by solving the
equations
\begin{eqnarray}
\mbox{Tr}\,(P) & = & 2[\cosh\frac{(a-b)}{2} + \cosh\frac{(a+b)}{2}]
\nonumber\\
\mbox{Tr}\,(P^2) & = & 2[\cosh(a-b) + \cosh(a+b)]
\label{abinp}
\end{eqnarray}
subject to the conditions on $a\/$ and $b\/$ appearing in
eq.~(\ref{invts-sp}).

Then we have the following convenient two-mode squeezing
transformation representing the equivalence class $(a,b)\/$:
\begin{eqnarray}
{\cal U}^{(0)}(a,b) & = & {\cal U}^{(0)}(a,0)\;{\cal U}^{(0)}(0,b),
\nonumber\\
{\cal U}^{(0)}(a,0) & = & \exp(iaK_2)\nonumber\\ & = &
\exp\left[\frac{-ia}{2}(q_1 p_1 + p_2 q_2)\right],\nonumber\\
{\cal U}^{(0)}(0,b) & = & \exp(ibL_1)\nonumber\\ & = &
\exp\left[\frac{ib}{2}(q_1 p_1 - q_2 p_2)\right].
\end{eqnarray}
The two factors ${\cal U}^{(0)}(a,0)\/$ and ${\cal U}^{(0)}(0,b)\/$
commute and may be written in either order, since according to eq.
(\ref{ncmp-comm}) the noncompact generators $K_2\/$ and $L_1\/$
commute.

Finally one can easily calculate the symplectic matrix $S^{(0)}(a,b)
\in Sp(4,\Re)\/$, corresponding to the metaplectic operator
${\cal U}^{(0)}(a,b)\/$, by using eq. (\ref{uopr-S}).  The result is:
\begin{eqnarray}
{\cal U}^{(0)}(a,b)^{-1}\;\xi\;{\cal U}^{(0)}(a,b) =
S^{(0)}(a,b)\;\xi,\nonumber\\ S^{(0)}(a,b) =
\mbox{diag.}\,\left(e^{(a-b)/2}, e^{(a+b)/2}, e^{-(a-b)/2},
e^{-(a+b)/2}\right).
\label{sab-diag}
\end{eqnarray}
Now we can clarify that the particular choice (\ref{kl-simple}) was
dictated by the desire to have $S^{(0)}(a,b)\/$ diagonal. This
element of $Sp(4,\Re)\/$ describes independent reciprocal scalings of
the standard quadrature components of each mode.  This amounts to
showing geometrically that it is possible to diagonalise every $P \in
\Pi$ using
conjugation by $U(2)$.

We illustrate our classification scheme of two-mode squeezing
transformations by giving two examples. The extensively studied
Caves-Schumaker~\cite{quadsq-one} transformation uses the operator
\begin{equation}
{\cal U}^{\mbox{(C-S)}}(z) = \exp\left(z\:a_1^\dagger\:a_2^\dagger -
z^\star\:a_1\:a_2\right).
\end{equation}
By appearance, this attempts to involve or entangle the two modes
maximally.  In our notation this squeezing transformation corresponds
to the generator combination
\begin{eqnarray}
z\:a_1^\dagger\:a_2^\dagger - z^\star\:a_1\:a_2 =
i(\vec{k}\cdot\vec{K} +
\vec{l}\cdot\vec{L}),\nonumber\\
\vec{k} = -2(0,0,\mbox{Im}\,z),\nonumber\\
\vec{l} = \phantom{-}2(0,0,\mbox{Re}\,z).
\end{eqnarray}
Thus the invariant parameters $a\/$ and $b\/$ have values
\begin{equation}
a = 2\vert z\vert,\quad b = 0.
\end{equation}
The Caves-Schumaker squeezing transformations, and their $U(2)\/$
conjugates, all taken together, form a one-parameter family or
one-dimensional line, in the $a-b\/$ octant. In that sense they are a
set of measure zero.

Another interesting case is a squeezing transformation that refers
essentially to a single mode but masquerades as a two-mode
transformation:
\begin{eqnarray}
{\cal U}(z;\alpha,\beta) = \exp\left[z(\alpha^\star a_1^\dagger +
\beta^\star a_2^\dagger)^2 - z^\star (\alpha a_1 + \beta a_2 )^2
\right],\nonumber\\
\vert\alpha\vert^2 + \vert\beta\vert^2 = 1.
\end{eqnarray}
After some simple algebra we find
\begin{eqnarray}
{\cal U}(z;\alpha,\beta) & = & \exp\left[i(\vec{k}\cdot\vec{K} +
\vec{l}\cdot\vec{L})\right],\nonumber\\
\vec{k} + i\vec{l} & = & 2z\left(-i(\alpha^\star{}^2 - \beta^\star{}^2),
(\alpha^\star{}^2 + \beta^\star{}^2), 2i\alpha^\star
\beta^\star\right).
\end{eqnarray}
The associated invariants and parameters are
\begin{eqnarray}
\Im_1 = 16\vert z\vert^4,\quad \Im_2 = 8\vert z\vert^2,\nonumber\\
a = b = 2\vert z\vert.
\end{eqnarray}
These equivalence classes thus lie along the line $a = b\/$ in the
octant, again a one-parameter family of zero measure. Our results are
depicted in Figure~\ref{ab-plane}.

We note here that the size of an equivalence class depends
sensitively upon the point $(a,b)$. Since for $a \neq 0\/,b \neq 0 $
and $a \neq b\/$ none of the generators of $U(2)$ or a linear
combination of them commute with $a K_2+b L_1$, we have a full four
parameter equivalence class.  For the cases $a \neq0, b=0\/$, and
$a=b$, respectively, the vanishing of the commutators $[J_2,K_2]$ and
$[Q+J_3,K_2+L_1]$ leads to reduction of the dimensionality of the
equivalence class to three.

We conclude this Section with a few comments. Each point $(a,b)\/$ in
the octant denotes an equivalence class of squeezing transformations,
whose dependences on $a\/$ and $b\/$ would be of physical
significance and would show up in a variety of $U(2)\/$-invariant
properties. The two-mode transformations so far discussed in the
literature lie basically along the two lines shown in
Figure~\ref{ab-plane}. In this sense, {\em most\/} of the
intrinsically distinct two-mode transformations, their effects on
various states, etc., remain to be explored. Those equivalence
classes $(a,b)\/$ for which $a > b\/$ involve the two modes in an
essential way. We may say purely qualitatively that the {\em
distance\/} of the point $(a,b)\/$ from the line $a=b\/$, or perhaps
better the expression $(1-b/a)\/$, is a measure of the extent to
which two independent modes are involved in the transformation. In
this sense, as remarked earlier, the Caves-Schumaker transformations
involve two modes maximally.

\section{Squeezed coherent and thermal states for two modes}
\label{four}
\setcounter{equation}{0}
The general two mode coherent state with complex two-component
displacement $\tilde{\alpha} = (\alpha_1 ,\alpha_2)\/$ is defined by
\begin{eqnarray}
\vert\tilde{\alpha}\rangle & = &
\exp\left(\tilde{\alpha}\cdot\tilde{a}^\dagger
- \tilde{\alpha}^\star\cdot\tilde{a}\right)\;\vert 0,0\rangle
\nonumber\\
& = & \exp\left(-\frac{1}{2}\vert\alpha_1\vert^2 - \frac{1}{2}\vert
\alpha_2\vert^2\right)\;\exp\left(\alpha_1 a_1^\dagger +
\alpha_2 a_2^\dagger \right)\;\vert 0,0\rangle.
\end{eqnarray}
For this state the means of the quadrature components $\xi_a\/$ do
not vanish in general:
\begin{equation}
\langle\tilde{\alpha}\vert\xi\vert\tilde{\alpha}\rangle =
\sqrt{2}\left(\mbox{Re}\,\alpha_1, \mbox{Re}\,
\alpha_2, \mbox{Im}\,\alpha_1,
\mbox{Im}\,\alpha_2\right)^T.
\end{equation}
The variance matrix is however independent of $\tilde{\alpha}\/$:
\begin{equation}
V\left(\vert\tilde{\alpha}\rangle\right) =
V\left(\vert\tilde{0}\rangle\right) = \frac{1}{2}\;{\bf 1}_{4\times
4}\,.
\end{equation}

The most general squeezed coherent state is obtained by applying
${\cal U}(P)\/$ for some $P\in\Pi\subset Sp(4,\Re)\/$ to $\vert
\tilde{\alpha}\rangle\/$ for some $\tilde{\alpha}\/$. This
${\cal U}(P)\/$ is conjugate, via some $U(2)\/$ element, to ${\cal
U}^{(0)}(a,b)\/$ for some $a,b\/$. Now the effect of a $U(2)\/$
transformation on $\vert\tilde{\alpha}\rangle\/$ is to give us
another coherent state $\vert\tilde{\alpha}^\prime\rangle\/$,
$\tilde{\alpha}^\prime\/$ being the $U(2)\/$ transform of
$\tilde{\alpha}\/$.  But the variance matrix is in any case
$\tilde{\alpha}$-independent. To examine the $U(2)\/$-invariant
squeezing condition, therefore, it suffices to examine the particular
class of squeezed coherent states
\begin{equation}
\vert\tilde{\alpha}; a, b\rangle = {\cal U}^{(0)}(a,b)\;\vert
\tilde{\alpha} \rangle.
\label{scs-u:vac}
\end{equation}
{}From eqs (\ref{vtransformation},\ref{sab-diag}), the calculation of
the variance matrix for this state is trivial, and it is in fact
diagonal:
\begin{eqnarray}
V\left(\vert\tilde{\alpha}; a, b\rangle\right) & = & S^{(0)}(a,b)\;
V\left(\vert\tilde{\alpha}\rangle\right)\; S^{(0)}(a,b)\nonumber\\ &
= & \frac{1}{2}\; S^{(0)}(2a,2b) \nonumber\\ & = & \frac{1}{2}\;
\mbox{diag.}\,\left(e^{(a-b)}, e^{(a+b)},
e^{(b-a)}, e^{-(a+b)}\right).
\end{eqnarray}
Since $a\/$ and $b\/$ are nonnegative, and in addition $a+b>0\/$, we
see that the least eigenvalue of this variance matrix is
\begin{equation}
\ell\left(V\left(\vert\tilde{\alpha}; a, b\rangle\right)\right) =
\frac{1}{2}\;e^{-(a + b)} < \frac{1}{2}.
\label{legvl}
\end{equation}
These states are thus always squeezed.

If we apply any passive $U(2)\/$ transformation $S(X,Y)\/$ to any one
of the states $\vert\tilde{\alpha}; a, b\rangle\/$ defined above, the
variance matrix will in general change as $V\rightarrow V^\prime =
S(X,Y)\;V\;S(X,Y)^T\/$; but its eigenvalue spectrum, and in
particular $\ell(V)\/$, remains unaltered. Thus all the states
symbolically written as ${\cal
U}(U(2))\,\vert\tilde{\alpha};a,b\rangle\/$, for various $U(2)\/$
elements, are squeezed to the same extent as
$\vert\tilde{\alpha};a,b\rangle\/$, and have $\ell(V)\/$ given by eq.
(\ref{legvl}).

The Schr\"{o}dinger wave functions for the subfamily of squeezed
coherent states (\ref{scs-u:vac}) are particularly simple, since they
are products of single mode squeezed coherent state wavefunctions:
\begin{eqnarray}
\langle q_1^\prime,q_2^\prime\vert\tilde{\alpha};a,b\rangle =
\psi^{(0)}(q_1^\prime;\alpha_1,a-b)\;
\psi^{(0)}(q_2^\prime;\alpha_2,a+b)\nonumber\\
\psi^{(0)}(q^\prime;\alpha,a) = \frac{e^{-a/4}}{\pi^{1/4}}\;\exp\left[
i\alpha\;\mbox{Im}\,\alpha - \frac{1}{2}\left( q^\prime\;e^{-a/2} -
\sqrt{2}\alpha\right)^2\right].
\end{eqnarray}
(For a general state ${\cal
U}(U(2))\,\vert\tilde{\alpha};a,b\rangle\/$, we do not expect such a
product form). When we set $b=0\/$ (Caves-Schumaker limit), both
factors show the same amount of squeezing; while when we set $a=b\/$
(essentially single mode situation) we see squeezing only in the
factor referring to the second mode. These features are as we would
have expected.

The next example we look at is the case of a two-mode thermal state
subjected to squeezing. The motivation in making this choice is that
the starting density operator is explicitly $U(2)\/$ invariant. The
normalized density operator corresponding to inverse temperature
$\beta = \hbar\omega/kT\/$ is described in the Fock and
Schr\"{o}dinger representations by:
\begin{mathletters}
\begin{eqnarray}
\rho_0(\beta) & = & (1 - e^{-\beta})^2\;\exp\left[-\beta(a_1^\dagger\:
a_1 + a_2^\dagger\:a_2)\right]\nonumber\\ & = & (1 -
e^{-\beta})^2\;\sum_{n_1,n_2=0}^{\infty} e^{-\beta (n_1 + n_2)}
\vert n_1,n_2\rangle\langle n_1,n_2\vert;\\
\rho_0(q_1^\prime,q_2^\prime,q_1^{\prime\prime},
q_2^{\prime\prime};\beta)
& = &
\frac{2}{\pi}\;\tanh^2\frac{\beta}{2}\;\exp\left[-\frac{1}{2}\left(
\tanh\frac{\beta}{2} + \coth\frac{\beta}{2}\right)\left(
q_1^{\prime}{}^2 + q_2^{\prime}{}^2 + q_1^{\prime\prime}{}^2 +
q_2^{\prime\prime}{}^2\right)\right. \nonumber\\ &&\mbox{}
-\left.\left(\tanh\frac{\beta}{2} - \coth\frac{\beta}{2}\right)
\left(q_1^{\prime}q_1^{\prime\prime} + q_2^{\prime}q_2^{\prime\prime}
\right) \right];
\end{eqnarray}
\end{mathletters}
with $U(2)\/$ invariance expressed by
\begin{eqnarray}
e^{\textstyle i\theta Q}\;\rho_0(\beta)\; e^{\textstyle -i\theta Q}&
=& e^{\textstyle i\vec{\alpha}\cdot\vec{J}} \;\rho_0(\beta)\;
e^{\textstyle -i\vec{\alpha}\cdot\vec{J}} = \rho_0(\beta).
\end{eqnarray}
Therefore it suffices to examine the properties of the density
operator obtained by conjugating $\rho_0(\beta)\/$ with ${\cal
U}^{(0)}(a,b)\/$:
\begin{equation}
\rho(\beta;a,b) = {\cal U}^{(0)}(a,b)\;\rho_0(\beta)\;
{\cal U}^{(0)}(a,b)^{-1}.
\label{rho-betaab}
\end{equation}
The most general squeezed thermal state is evidently
\begin{equation}
{\cal U}(U(2))\;\rho(\beta;a,b)\;{\cal U}(U(2))^{-1},
\end{equation}
but this has the same squeezing properties as $\rho(\beta;a,b)\/$.

For the thermal state $\rho_0(\beta)\/$ the variance matrix is well
known~\cite{multimode}:
\begin{equation}
V\left(\rho_0(\beta)\right) = \frac{1}{2}\;\coth\frac{\beta}{2}\;
{\bf 1}_{4\times 4}
\end{equation}
Therefore for the particular set of squeezed thermal states
(\ref{rho-betaab}), we have diagonal variance matrices:
\begin{eqnarray}
V\left(\rho(\beta;a,b)\right) & = &
S^{(0)}(a,b)\;V\left(\rho_0(\beta)
\right)\; S^{(0)}(a,b)^T\nonumber\\
& = &\frac{1}{2}\;\coth\frac{\beta}{2}\;S^{(0)}(2a,2b)\nonumber\\ & =
&\frac{1}{2}\;\coth\frac{\beta}{2}\;\mbox{diag.}\,
\left(e^{(a-b)}, e^{(a+b)}, e^{(b-a)}, e^{-a-b}\right).
\end{eqnarray}
The least eigenvalue is evidently
\begin{equation}
\ell(V) =  \frac{1}{2}\;\coth\frac{\beta}{2}\;e^{-(a+b)},
\end{equation}
so for a given temperature, squeezing sets in when
\begin{equation}
a + b > \ln\coth\frac{\beta}{2}.
\end{equation}
In Figure~\ref{ab-plane}, this region consists of all points in the
$a-b\/$ octant {\em to the right of\/} the line $a + b =
\ln\coth\beta/2\/$, which is a line perpendicular to the line
$a=b\/$ and at a distance $\ln\coth(\beta/2)/\sqrt{2}\/$ from the
origin.

\section{ Detection schemes and role of $U(2)$ transformations}
\label{five}
\setcounter{equation}{0}
We have so far not specified in any detail the two orthogonal modes
of radiation being subjected to squeezing. Let us at this point
consider a situation well studied experimentally by the heterodyne
detection scheme~\cite{heterodyne}.  Here the two modes differ only
slightly in frequency, but are otherwise similar. In this kind of
experimental arrangement, what is actually measured is the
fluctuation of a certain photocurrent, and this in turn gives the
fluctuation or variance of the $q\/$-quadrature component of a
particular (passive) combination of the original modes. The
combinations of $a_1\/$ and $a_2\/$ that are involved form the
one-parameter family
\begin{equation}
a(\psi) = \frac{1}{\sqrt{2}}\;(a_1 + a_2)\;e^{-i\psi/2}\;,\quad 0\leq
\psi<4\pi.
\end{equation}
This can be regarded as the first component of a $U(2)\/$-transformed
pair $a_1^\prime\/$, $a_2^\prime\/$:
\begin{eqnarray}
\left(\begin{array}{c} a_1^\prime \\ a_2^\prime\end{array}\right)
& = & \frac{1}{\sqrt{2}}
\left(\begin{array}{rl} e^{-i\psi/2} & e^{-i\psi/2} \\
-e^{i\psi/2} & e^{i\psi/2}\end{array}\right)
\left(\begin{array}{c} a_1 \\ a_2\end{array}\right),\nonumber\\
a_1^\prime & \equiv & a(\psi).
\label{het-utwo}
\end{eqnarray}
The hermitian quadrature component whose fluctuation is measured is
\begin{eqnarray}
q(\psi) & = &\frac{1}{\sqrt{2}}\;\left( a(\psi) +
a(\psi)^\dagger\right)
\nonumber\\
& = & \frac{1}{\sqrt{2}}\;\left( q_1 + q_2\right)\;\cos\frac{\psi}{2}
+
\frac{1}{\sqrt{2}}\;\left( p_1 + p_2\right)\;\sin\frac{\psi}{2}.
\end{eqnarray}
The only experimentally adjustable parameter here is the angle
$\psi\/$.  The family of $U(2)\/$ elements realised in the heterodyne
scheme is thus only the one-parameter set given in eq.
(\ref{het-utwo}) parametrised by $\psi\/$, and belonging to
$SU(2)\/$:
\begin{equation}
U_H(\psi) = \frac{1}{\sqrt{2}}\left(\begin{array}{rl} e^{-i\psi/2} &
e^{-i\psi/2} \\ -e^{i\psi/2} & e^{i\psi/2}\end{array}\right) \in
SU(2).
\end{equation}
We notice that this is {\em not\/} a one-parameter subgroup of
$SU(2)\/$; in particular even the identity element of the group is
not contained here.

With this description of the heterodyne setup in our framework, let
us see to what extent it can be used to detect $U(2)\/$-invariant
squeezing. Now a general two-mode state $\rho\/$ with variance matrix
$V\/$, even if it is squeezed in the intrinsic sense of
eq.~(\ref{sq-crt-mat}), may not be manifestly squeezed. That is, it
may happen that $V_{aa}\geq 1/2\/$ for all $a=1,\ldots ,4\/$. As our
discussion in Section~(\ref{two}) shows, we need to be able to {\em
experimentally\/} realise a general $U(2)\/$-transformation applied
to the state $\rho\/$, and change its variance matrix to a form where
one of its diagonal entries (say the leading one) becomes less than
1/2. However the heterodyne method is generally unable to do this job
for us, as it can only realise the one-parameter subset of $SU(2)\/$
transformations $U_H(\psi)\/$ for $0\leq \psi<4\pi\/$.

In the two examples of squeezed coherent states and squeezed thermal
states studied in the previous Section, we have a family of states
related to each other by conjugation with $U(2)$ for each point in
the $a-b\/$ plane. Each equivalence class has appropriate
dimensionality depending upon the point $(a,b)$ as explained in
Section~\ref{three}.  It turns out that for each $(a,b)\/$ the
heterodyning scheme can detect squeezing in only a one parameter
subset of the family of states.  Although heterodyning detection
covers the whole $a-b\/$ plane it does not reach all the states
corresponding to each point in the $a-b\/$ plane.  For example, in
the representative chosen in eqs (\ref{scs-u:vac},
\ref{rho-betaab}), for which the variance matrix is already
diagonal, the squeezing cannot be detected by this scheme because of
the absence of the identity in $U_H$!  It should be possible to
detect squeezing in these states by a suitably modified scheme.  We
wish to emphasize that there is a definite need to be able to
experimentally implement the most general elementof $U(2)\/$. This
would allow the experimenter to detect the degree of squeezing
unambiguously, if the state is squeezed, without any prior knowledge
of the elements of the initial variance matrix.

Having stressed the need to implement arbitrary $U(2)$
transformations on the two modes of radiation in order to reach the
proper quadrature to exhibit squeezing, we now describe how it can be
achieved in some situations. We discuss two particular cases of the
two modes involved, the first when the two modes have the same
frequency but different directions of propagation, and the second
when the modes have the same frequency and direction of propagation
but different polarisations.

The experimental setup for the first case is shown in
Figure~\ref{mach-zehnder}. We achieve an arbitrary $U(2)$
transformation on the two modes by using a Mach-Zehnder
interferometer with two 50:50 beam splitters (BS$_1$ and BS$_2$) and
appropriate phase shifters~\cite{mach-zehnder-p}.  The input modes
with annihilation operators $a_1$ and $a_2$ are subjected to equal
and opposite phase shifts by angles $\phi$ and $-\phi$, then the
modes are mixed in the beam splitter BS$_1$, the mixed modes again
undergo equal and opposite phase shifts by angles $\theta$ and
$-\theta$ and further mixing through the beam splitter BS$_2$.
Finally they undergo unequal phase shifts by angles $\psi_1$ and
$\psi_2$.  If the annihilation operators at the output are
$a_1^\prime$ and $a_2^\prime$ then all the above operations when
combined are implemented through the transformation
\begin{equation}
\left(\begin{array}{c} a_1^\prime \\ a_2^\prime\end{array}\right)=
\left(\begin{array}{rl} e^{i(\phi+\psi_1)}\cos{\theta} &
-ie^{-i(\phi-\psi_1)}\sin{\theta} \\ -ie^{i(\phi+\psi_2)}\sin{\theta}
& e^{-i(\phi-\psi_2)}\cos{\theta}\end{array}\right)
\left(\begin{array}{c} a_1 \\ a_2\end{array}\right)
\end{equation}
relating the two sets of annihilation operators.  The above matrix is
the most general $U(2)$ transformation matrix. We note here that if $
\psi_2=-\psi_1$ then the
transformation matrix is the most general $SU(2)$ transformation.
Going from $SU(2)$ to $U(2)$ is just a matter of overall phase and
can also be achieved by free propagation.

For the second case when the two modes differ only in polarisation we
achieve the arbitrary $U(2)$ transformation by using two quarter wave
plates (Q$_1$, Q$_2$) and a half wave plate H as shown in
Figure~\ref{polarisation}. The detailed discussion of this set up is
given in~\cite{polarisation-p}.  It turns out that the configuration
Q-H-Q is not the only one but Q-Q-H and H-Q-Q also accomplish the
same result, as shown in~\cite{polarisation-p}.  We basically have
three elements, a quarter wave plate Q$_1$, a half wave plate H and a
quarter wave plate Q$_2$ all three of them being coaxially mounted
and with their slow axes in the $x-y$ plane making angles of $\alpha,
\beta$ and
$\gamma$ respectively with the $\hat{x}$ axis. The two modes having
annihilation operators $a_1$ and $a_2$ moving along the $\hat{z}$
direction pass through this arrangement. If the annihilation
operators at the output are $a_1^\prime$ and $a_2^\prime$ then they
are related to the operators at the input by an $SU(2)$
transformation given in terms of $\alpha,
\beta$ and $\gamma$. By changing these parameters one
can reproduce any desired $SU(2)$ element. As has been pointed out
earlier going to $U(2)$ now is just a matter of free propagation.

In both the above cases, by going to the proper $U(2)\/$ element we
can make the squeezing(if present) manifest and bring it to the
leading diagonal element of the variance matrix i.e. in the
quadrature $q_1^\prime=\frac{1}{\sqrt{2}}(a_1^{\prime}+a_1^{\dagger
\prime})$.
The squeezing in this quadrature can now be measured by any standard
one mode detection method.

These remarks show on the one hand the way the heterodyning scheme
fits into our general analysis, and on the other hand the need to
devise new schemes capable of realising all elements of $U(2)\/$,
tailored to the definition of the two modes involved.

\section{Concluding remarks}
\label{concluding}
We have presented a classification scheme for two-mode squeezing
transformations, based on the structure of the real four dimensional
symplectic group $Sp(4,\Re)\/$, and the separation of its elements
into passive (compact) and active (noncompact) types. The structure
and action of the maximal compact subgroup $U(2)\/$ in $Sp(4,\Re)\/$,
and the $U(n)\/$-invariant squeezing criterion formulated elsewhere
for a general $n\/$-mode system, have guided our considerations. All
our work is in the metaplectic unitary representation of
$Sp(4,\Re)\/$; and the local isomorphism $Sp(4,\Re) \approx
SO(3,2)\/$ has led to a convenient geometric picturization of the
situation.

As emphasized in Section~\ref{three} the squeezing transformations
${\cal U}(P)\/$, $P\in \Pi\subset Sp(4,\Re)\/$, do not form a
subgroup of $Sp(4,\Re)\/$. The breakup of these transformations into
equivalence classes, based on the effect of conjugation by elements
of $U(2)\/$ is the only natural available classification procedure.
This is because the definition of equivalence classes for any set of
objects has to be based on an equivalence relation defined on that
set. Thus, we have treated two elements $P\/$, $P^\prime\in \Pi\/$ as
intrinsically equivalent if
\begin{equation}
P^\prime=S(X,Y)\;P\;S(X,Y)^T \quad \mbox{for some}\quad X - iY \in
U(2).
\end{equation}
It should however be realised that the detailed effects of action by
${\cal U}(P)\/$ and ${\cal U}(P^\prime)\/$ on a {\em general\/}
initial two mode state $\rho_0\/$, as seen in the changes caused in
the variance matrix $V(\rho_0)\/$, need not be identical. Since this
is a subtle and important point we spell it out in detail. Starting
from a general state $\rho_0\/$, action by a squeezing transformation
leads to a new state
\begin{equation}
\rho = {\cal U}(P) \;\rho_0\;{\cal U}(P)^{-1}.
\end{equation}
As seen in Section~\ref{three}, any ${\cal U}(P)\/$ is expressible in
terms of a representative element ${\cal U}^{(0)}(a,b)\/$ as
\begin{equation}
{\cal U}(P) = {\cal U}(S(X,Y))\;{\cal U}^{(0)}(a,b)\; {\cal
U}(S(X,Y))^{-1}
\end{equation}
for suitable $X-iY\in U(2)\/$. Therefore we have
\begin{equation}
\rho  = {\cal U}(S(X,Y))\;{\cal U}^{(0)}(a,b)\;
{\cal U}(S(X,Y))^{-1}\;\rho_0\; {\cal U}(S(X,Y))\;{\cal
U}^{(0)}(a,b)^{-1}\; {\cal U}(S(X,Y))^{-1}.
\end{equation}
This leads by eq. (\ref{vtransformation}) to the relation
\begin{equation}
V(\rho) = S(X,Y)\;S^{(0)}(a,b)\;S(X,Y)^{T}\;V(\rho_0)\;
S(X,Y)\;S^{(0)}(a,b)\;S(X,Y)^{T}
\end{equation}
between the two variance matrices. Now the first and last factors on
the right hand side here have no influence on the spectrum, and so on
the least eigenvalue, of $V(\rho)\/$. Therefore the squeezed or
nonsqueezed nature of $\rho\/$ is actually determined by the least
eigenvalue of the matrix
\begin{equation}
S(X,Y)^{T}\;V(\rho)\;S(X,Y) = S^{(0)}(a,b)\;S(X,Y)^{T}\;V(\rho_0)\;
S(X,Y) \;S^{(0)}(a,b).
\label{vgeneral}
\end{equation}
But now the right hand side is in {\em general\/} dependent not only
on the invariant parameters $a,b\/$ but also on $X,Y\/$. In the
examples studied in Section~\ref{four}, namely where $\rho_0\/$ is a
coherent state or an isotropic thermal state, $V(\rho_0)\/$ happens
to be a multiple of the identity matrix, so that on the right hand
side of eq. (\ref{vgeneral}) the dependence on $X,Y\/$ cancels. But
this need not happen in general. Thus for instance if we take for
$\rho_0\/$ an anisotropic thermal state with unequal temperatures for
the two modes, we have only $U(1)\times U(1)\/$, rather than
$U(2)\/$, invariance for this $\rho_0\/$; so the least eigenvalue
$\ell (V(\rho))\/$ of $V(\rho)\/$ will depend on $a,b\/$ and on two
out of the four $U(2)\/$ parameters present in $X - iY\/$.  One can
easily convince oneself that the only situation where
$S(X,Y)^T\;V(\rho_0)\;S(X,Y) = V(\rho_0)\/$ independent of $X\/$ and
$Y\/$ is when $V(\rho_0)\/$ is a multiple of the unit matrix; and the
isotropic thermal states do reproduce all such cases.  Therefore a
more detailed study of the effect of a general squeezing
transformation on initial states $\rho_0\/$ with nontrivial
$V(\rho_0)\/$ is of considerable interest.

A related important question is the following: Let us take two
squeezing transformations $P_1, P_2\in \Pi\/$ belonging to
equivalence classes $(a_1,b_1)$, $(a_2,b_2)\/$ respectively, which
could coincide. The product $P_1 P_2\/$ will in general be of the
form $S(X,Y)\:P\/$ with $S(X,Y)\in U(2)\/$ and $P\in \Pi$.  If $P\/$
belongs to the equivalence class $(a,b)\/$ we wish to determine this
class in terms of $P_1\/$ and $P_2$.  Using the fact
$Tr(P^2)=Tr((P_1P_2)(P_1P_2)^T)\/$ and eq. (\ref{abinp}) we arrive at
the following relations
\begin{eqnarray}
2[\cosh{(a-b)}+\cosh{(a+b)}]=Tr((P_1P_2)(P_1P_2)^T)
\nonumber \\
2[\cosh{2(a-b)}+\cosh{2(a+b)}]=Tr(((P_1P_2)(P_1P_2)^T)^2)
\end{eqnarray}
which can be solved to find $(a,b)\/$.  We note here that $(a,b)$ do
depend not only upon $(a_1,b_1)$ and $(a_2,b_2)$ but also on the
actual elements chosen from each of these classes. So we do not have
a notion of class multiplication among these equivalence classes.

Finally we call attention to our considerations in Section~\ref{five}
and to the need for being able to experimentally implement or realize
general passive elements of the subgroup $U(2)\/$ of $Sp(4,\Re)\/$
for each given choice of the independent modes in a two-mode system.
Once this is achieved, for any given state we can bring out in an
explicit or manifest fashion its squeezing nature (provided it is
squeezed) by altering its variance matrix and making the least
eigenvalue appear in the leading position on the diagonal. This also
means that we would be able to experimentally measure the fluctuation
in the quadrature variable isolating the least eigenvalue. As
extensively discussed elsewhere, these considerations which exploit
the richness of the geometry underlying the symplectic group do not
require complete diagonalization of the variance matrix at
all~\cite{multimode}.

In the following paper we shall examine $U(2)\/$-invariant properties
of two mode squeezed states which go beyond the level of second order
moments of the quadrature operators.  

\newpage
\setcounter{figure}{0}
\setcounter{equation}{0}
\begin{figure}
\setlength{\unitlength}{1cm}
\begin{picture}(15,15)(0,0)
\thicklines
\put(3,2){\line(1,0){10}}
\put(3,2){\line(0,1){10}}
\put(3,2){\line(1,1){9}}
\put(2.5,1.5){(0,0)}
\put(6.5,1.5){$a$}
\put(7,1.6){\vector(1,0){1}}
\put(2.5,5.5){$b$}
\put(2.6,6.0){\vector(0,1){1}}
\put(7,2){\line(-1,1){2}}
\put(6.5,4.5){Squeezed thermal region}
\put(7,2.25){C-S transformations $(b=0)$}
\put(4.6,8.0){\shortstack[l]{Essentially single \\mode
transformations}}
\put(7.9,8.5){$(a=b)$}
\put(4.5,2.3){\shortstack[l]{$a+b=$ \\ $\ln\coth\frac{\beta}{2}$}}
\end{picture}
\caption{Equivalence classes of two-mode squeezing transformations,
Caves-Schumaker (C-S) and single mode limits, squeezed thermal
region.}
\label{ab-plane}
\end{figure}
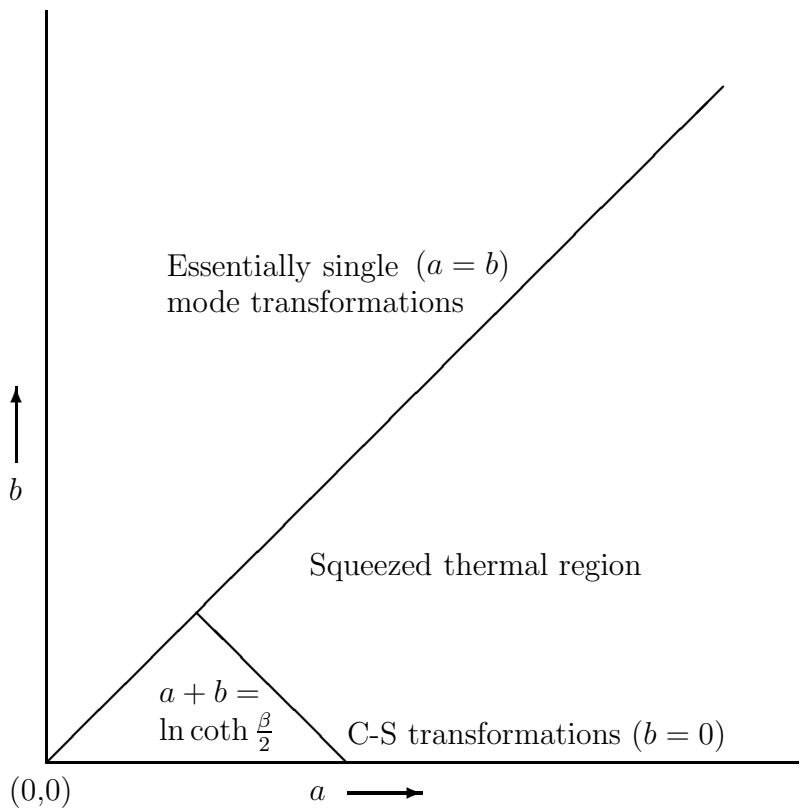
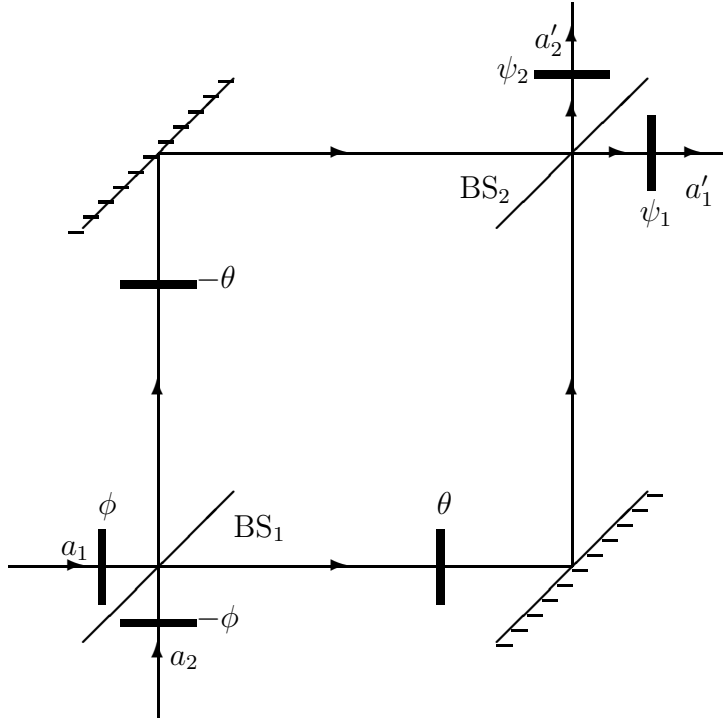
\begin{figure}
\setlength{\unitlength}{1cm}
\begin{picture}(15,15)(0,0)
\thicklines
\put(2,4){\line(1,0){7.5}}
\put(4,9.5){\line(1,0){7.5}}
\put(4,2){\line(0,1){7.5}}
\put(9.5,4){\line(0,1){7.5}}
\put(3,3){\line(1,1){2}}
\put(3,8.5){\line(1,1){2}}
\put(8.5,3){\line(1,1){2}}
\put(8.5,8.5){\line(1,1){2}}
\multiput(2.8,8.45)(0.2,0.2){11}{\line(1,0){0.2}}
\multiput(8.5,2.95)(0.2,0.2){11}{\line(1,0){0.2}}
\put(3,4.02){\vector(1,0){0}}
\put(6.5,4.02){\vector(1,0){0}}
\put(6.5,9.52){\vector(1,0){0}}
\multiput(10.2,9.52)(1,0){2}{\vector(1,0){0}}
\put(3.981,3){\vector(0,1){0}}
\put(3.981,6.5){\vector(0,1){0}}
\put(9.486,6.5){\vector(0,1){0}}
\multiput(9.486,10.2)(0,1){2}{\vector(0,1){0}}
\put(2.7,4.15){$a_1$}
\put(4.15,2.7){$a_2$}
\put(11,8.9){$a_1^\prime$}
\put(9,10.9){$a_2^\prime$}
\put(5,4.4){$\mbox{BS}_1$}
\put(8,8.9){$\mbox{BS}_2$}
\multiput(3.2,3.5)(4.5,0){2}{\rule{1mm}{1cm}}
\put(3.2,4.7){$\phi$}
\put(7.7,4.7){$\theta$}
\multiput(3.5,3.2)(0,4.5){2}{\rule{1cm}{1mm}}
\put(4.5,3.2){$- \phi$}
\put(4.5,7.7){$- \theta$}
\put(10.5,9.0){\rule{1mm}{1cm}}
\put(9.0,10.5){\rule{1cm}{1mm}}
\put(8.5,10.5){$\psi_2$}
\put(10.4,8.6){$\psi_1$}
\end{picture}
\caption{Mach-Zehnder Interferometer implementing arbitrary
$U(2)$ transformation on two modes at the same frequency but
differing in their direction of propagation. BS$_1$ and BS$_2$ are
50:50 beam splitters and thick lines are phase shifters by angles
indicated. $a_1$, $a_2$ are the annihilation operators at the input
port and $a_1^\prime$, $a_2^\prime$ are the annihilation operators at
the output port}
\label{mach-zehnder}
\end{figure}
\begin{figure}
\setlength{\unitlength}{1cm}
\begin{picture}(15,15)(0,0)
\thicklines
\put(2,7){\line(1,0){9}}
\multiput(3,7.02)(2.4,0){4}{\vector(1,0){0}}
\multiput(3.3,7.02)(2.4,0){4}{\vector(1,0){0}}
\multiput(4.2,6.3)(4.8,0){2}{\rule{1mm}{1.4cm}}
\put(6.6,6.4){\fbox{\rule{0mm}{1.2cm}}}
\put(2,8){\line(1,0){1}}
\put(2.95,8){\vector(1,0){0}}
\put(3.1,7.9){$\hat{z}$}
\put(2,9.05){\vector(0,1){0}}
\put(2,8){\line(0,1){1}}
\put(2.5,8.55){\vector(1,1){0}}
\put(1.9,9.1){$\hat{x}$}
\put(2,8){\line(1,1){0.5}}
\put(2.6,8.6){$\hat{y}$}
\put(2.7,6.4){$a_1 a_2$}
\put(10,6.4){$a_1^\prime a_2^\prime$}
\put(4.1,8){$\mbox{Q}_1$}
\put(8.95,8){$\mbox{Q}_2$}
\put(6.6,8){H}
\put(3.8,7.2){$ \alpha $}
\put(8.65,7.2){$ \gamma $}
\put(6.2,7.2){$ \beta $}
\end{picture}
\caption{Implementation of arbitrary $SU(2)$ element on two modes
with the same frequency and directions of propagation, but different
polarisations. Here Q$_1$, Q$_2$ are quarter wave plates and H is the
half wave plate. $\alpha,\beta, \mbox{and} \gamma$ are the
angles which the slow axes of Q$_1$, H and Q$_2$ make with the
$\hat{x}$ axis respectively. $a_1$, $a_2$ are the
annihilation operators at the input port and $a_1^\prime$,
$a_2^\prime$ are the annihilation operators at the output port }
\label{polarisation}
\end{figure}
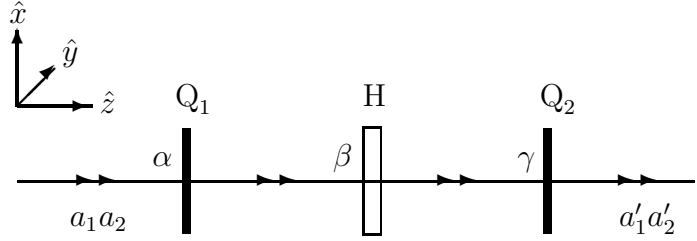

\begin{references}
\bibitem[\star]{email} email arvind@physics.iisc.ernet.in
\bibitem[\dag]{jncasr} Also at Jawaharlal Nehru Centre for
Advanced Scientific
Research, Indian Institute of Science Campus, Bangalore - 560 012,
India.
\bibitem{discovery} D.~Stoler, Physical Review D {\bf 1}, 3217 (1970);
H.~P.~Yuen, Physical Review A {\bf 13}, 2226 (1976);
D.~F.~Walls, Nature {\bf 306}, 141 (1983);
J.~N.~Hollenhorst, Physical Review D {\bf 19}, 1669 (1979).
\bibitem{expt-one} R.~E.~Slusher, L.~W.~Hollberg,
B.~Yurke, J.~C.~Mertz,
and J.~F.~Valley,  Physical Review Letters {\bf 55}, 2409 (1985);
R.~M.~Shelby, M.~D.~Levenson, S.~H.~Perlmutter, R.~G.~DeVoe,
and D.~F.~Walls,
Physical Review Letters {\bf 57}, 691 (1986);
Ling-An Wu, J.~H.~Kimble, J.~L.~Hall, and H.~Wu,
Physical Review Letters
{\bf 57}, 2520 (1986); M.~W.~Maeda, P.~Kumar, and J.~H.~Shapiro,
Optics Letters {\bf 12}, 161 (1987).
\bibitem{expt-two} B.~L.~Schumaker, S.~H.~Perlmutter, R.~M.~Shelby and
M.~D.~Levenson,  Physical Review Letters {\bf 58}, 357 (1987);
M.~G.~Raizen, L.~A.~Orosco, M.~Xiao, T.~L.~Boyd and H.~J.~Kimble,
Physical Review Letters {\bf 59}, 198 (1987); A.~Heidmann,
R.~J.~Horowicz, S.~Reynaud, E.~Giacobino, C.~Fabre, and G.~Camy,
Physical Review Letters {\bf 59}, 2555 (1987);
Z.~Y.~Ou, S.~F.~Pereira, H.~J.~Kimble and
K.~C.~Peng,  Physical Review Letters {\bf 68}, 3663 (1992).
\bibitem{expt-three} P.~Kumar, O.~Aytur and J.~Huang,
Physical Review Letters {\bf 64}, 1015 (1990);
A.~Sizmann, R.~J.~Horowicz, E.~Wagner and G.~Leuchs, Optics
Communications
{\bf 80}, 138 (1990);
M.~Rosenbluh and R.~M.~Shelby,  Physical Review Letters
{\bf 66}, 153 (1991);
K. Bergman and H.~A.~Haus, Optics Letters {\bf 16}, 663 (1991);
E.~S.~Polzik, R.~C.~Carri and H.~J.~Kimble, Physical Review Letters
{\bf 68}, 3020 (1992).
\bibitem{reviews} For recent reviews, see, R.~Loudon and P.~L.~Knight,
Journal of Modern Optics {\bf 34}, 709 (1987); M.~C.~Teich
and B.~E.~A.~Saleh,
Quatum Optics {\bf 1}, 153 (1990), S.~Reynaud, A.~Heidmann,
E.~Giacobino and C.~Fabre, in {\it Progress in Optics\/}, Vol. 30,
ed. E.~Wolf (North-Holland, Amsterdam, 1992);
C. Fabre, Physics Reports {\bf 219}, 215 (1992); H.~J.~Kimble,
Physics Reports {\bf 219}, 227 (1992).
\bibitem{quadsq-one} C.~M.~Caves and B.~L.~Schumaker,
Physical Review A {\bf 31} , 3068 (1985);
B.~L.~Schumaker and C.~M.~Caves, Physical Review A
{\bf 31} , 3093 (1985);
B.~L.~Schumaker, Physics Reports {\bf 135}, 317 (1986);
C.~M.~Caves, Physical Review D {\bf 23}, 1693 (1981).
\bibitem{pcs} G. S. Agarwal,  Physical Review Letters
{\bf 57}, 827 (1986);
G. S. Agarwal, Journal of the Optical Society of America B
{\bf 5}, 1940 (1988).
\bibitem{pnd-two-mode} A.~K.~Eckert and P.~L.~Knight,
American Journal Physics {\bf 57}, 692 (1989); C.~M.~Caves,
C.~Zhu, G.~J.~Milburn and W.~Schleich, Physical Review A
{\bf 43}, 3854 (1991);
M.~Selvadoray, M.~Sanjay Kumar and R.~Simon, preprint
I.~M.~Sc. - 93/22 (1993).
\bibitem{high-order} C.~K.~Hong and L.~Mandel,  Physical Review Letters
{\bf 54}, 323 (1985); Physical Review A {\bf 32}, 974 (1985);
C.~C.~Gerry and P.~J.~Moyer, Physical Review A {\bf 38}, 5665 (1988);
M.~S.~Kim, F.~A.~M.~de~Oliviera and P.~L.~Knight, Physical Review A
{\bf 40}, 2494 (1989); J.~J.~Gong and P.~K.~Aravind, Physical Review A
{\bf 46}, 1586 (1992);
M.~Hillery, Physical Review A {\bf 36}, 3796 (1987){}{}{}.
\bibitem{multimode} R.~Simon, N.~Mukunda and B.~Dutta,
Physical Review A {\bf 49}, 1567 (1994).
\bibitem{stone-neumann} T.~F.~Jordan, {\it Linear operators in
quantum mechanics\/}, (John Wiley, New York, 1974);
G.~Lion and M.~Vergne, {\it The Weil Representation,
Maslov Index, and Theta series\/}, (Birkhauser, Basel, 1980).
\bibitem{metaplectic} A.~Weil, Acta Mathematica {\bf 111}, 143 (1964);
V.~Guillemin and S.~Sternberg, {\it Symplectic Techniques
in Physics\/}, Cambridge University Press, (1984);
R.~Simon and N.~Mukunda, {\it The two-dimensional symplectic and
metaplectic groups and their universal cover\/} in {\it Symmetries
in Science \uppercase{vi}: From the Rotation
Group to Quantum Algebras\/}, B.~Gruber (ed.), Plenum Press
(to appear).
\bibitem{uone-sqcrt} N.~Mukunda, Current Science {\bf 59\/},
1135 (1990); B.~Dutta, N.~Mukunda, R.~Simon and A.~Subramaniam,
Journal of the Optical Society of America B {\bf 10}, 253 (1993).
\bibitem{sp4r-so32} The connection between the groups $Sp(4,\Re)\/$
and $SO(3,2)\/$ has been noticed and exploited in many physical
contexts for a very long time. These include the infinite component
relativistic wave equations constructed by Majorana, the new Dirac
equation for a relativistic point particle with two internal harmonic
oscillator degrees of freedom, the theory of relativistic objects
with internal structure and classification of anisotropic Schell
model beams in optics. See E. Majorana, Nuovo Cimento {\bf 9}, 335
(1932); P.~A.~M.~Dirac, Journal of Mathematical Physics {\bf 4}, 901
(1963); N.~Mukunda, Physical Review {\bf 183}, 1486 (1969);
E.~C.~G.~Sudarshan and N.~Mukunda, Physical Review D {\bf 1}, 576
(1970); P.~A.~M.~Dirac, Proceedings of the Royal Society A {\bf 322},
435 (1971); {\it ibid\/} A {\bf 328}, 1 (1972); N.~Mukunda, H.~van
Dam and L.~C.~Biedenharn, {\it Relativistic Models of Extended
Hadrons obeying a mass spin trajectory\/}, Lecture Notes in Physics,
Vol. 165, Springer-Verlag 1982; R.~Simon, E.~C.~G.~Sudarshan and
N.~Mukunda, Physical Review A {\bf 31}, 2419 (1985); Physical Review
A {\bf 36}, 3868 (1987).  A recent paper exploiting these ideas for
squeezing problems is D.~Han, Y.~S.~Kim, M.~E.~Noz and L. Yeh,
Journal of Mathematical Physics {\bf 34}, 5493 (1993).
\bibitem{gilmore} R. Gilmore, {\it Lie Groups, Lie Algebras and some of
their applications\/}, John Wiley and Sons, New York USA(1974)
\bibitem{siegel} C. L. Siegel, American Journal of Mathematics
{\bf \uppercase{lxv}}, 1 (1943), reprinted as C. L. Siegel {\it
Symplectic Geometry\/}, Academic Press (New York) 1964.
\bibitem{polar-decomposition} Arvind, B.~Dutta, N.~Mukunda and
R.~Simon, {\it The real symplectic groups in quantum mechanics and
optics\/}, IISc preprint in preparation.
\bibitem{wigner} E. P. Wigner, {\it Symmetries and Reflections},
Indiana University Press, 1967.
\bibitem{heterodyne} H.~P.~Yuen and J.~H.~Shapiro, IEEE Transactions
on Information Theory {\bf 24}, 657 (1978); M.~D.~Levenson,
R.~M.~Shelby, A.~Aspect, M.~Reid and D.~F.~Walls, Physical Review A
{\bf 32}, 1550 (1985); M.~J.~Collett, R.~Loudon and C.~W.~Gardiner,
Journal of Modern Optics {\bf 34}, 881 (1987).
\bibitem{mach-zehnder-p} B.~Yurke, S.~L.~McCall and J.~R.~Klauder,
Physical Review A {\bf 33}, 4033 (1986); R.~A.~Campos, B.~E.~A.~Saleh
and M.~C.~Teich, Physical Review A {\bf 40}, 1371 (1989);
R.~F.~Bishop and A.~Vourdas, Zeitschrift f\"{u}r Phyzik {\bf 71}, 527
(1988); C.~T.~Lee, Physical Review A {\bf 42}, 4193 (1990); D.~Han,
Y.~S.~Kim and M.~E.~Noz, Physical Review A {\bf 41} , 6233 (1990).
\bibitem{polarisation-p} R.~Simon and N.~Mukunda,
Physics Letters A {\bf 143}, 165 (1990);
\end{references}
\end{document}